# Membrane Nanodomains Homeostasis During Propofol Anesthesia as Function of Dosage and Temperature

Weixiang Jin, Michael Zucker, and Arnd Pralle*

Dept. of Physics, University at Buffalo, SUNY, Buffalo, NY 14260-1500, USA

*Email: apralle@buffalo.edu

**Abstract**

Some anesthetics bind and potentiate γ-aminobutyric-acid-type receptors, but no universal mechanism for general anesthesia is known. Furthermore, often encountered complications such as anesthesia induced amnesia are not understood. General anesthetics are hydrophobic molecules easily dissolving into lipid bilayers. Recently, it was shown that general anesthetics perturb phase separation in vesicles extracted from fixed cells. Unclear is whether under physiological conditions general anesthetics induce perturbation of the lipid bilayer, and whether this contributes to the transient loss of consciousness or anesthesia side effects. Here we show that propofol perturbs lipid nanodomains in the outer and inner leaflet of the plasma membrane in intact cells, affecting membrane nanodomains in a concentration dependent manner: 1 µM to 5 µM propofol destabilize nanodomains; however, propofol concentrations higher than 5 µM stabilize nanodomains with time. Stabilization occurs only at physiological temperature and in intact cells. This process requires ARP2/3 mediated actin nucleation and Myosin II activity. The rate of nanodomain stabilization is potentiated by GABA receptor activity. Our results show that active nanodomain homeostasis counteracts the initial disruption causing large changes in cortical actin.

**Significance Statement**

General anesthesia is a routine medical procedure with few complications, yet a small number of patients experience side-effects that persist for weeks and months. Very young children are at risk for effects on brain development. Elderly patients often exhibit subsequent amnesia. Here, we show that the general anesthetic propofol perturbs the ultrastructure of the lipid bilayer of the cell membrane in intact cells. Initially propofol destabilized lipid nanodomains. However, with increasing incubation time and propofol concentration, the effect is reversed and nanodomains are further stabilized. We show that this stabilization is caused by the activation of the actin cortex under the membrane. These perturbations of membrane bilayer and cortical actin may explain how propofol affects neuronal plasticity at synapses.

**Introduction**

General anesthesia is performed more than 60,000 times each day across the USA, and 235 million times around the globe. A portion of patients exhibit subsequent cognitive impairment, including memory deficits, after undergoing anesthesia.(Bortolon, Weglinski and Sprung, 2005; Han *et al.*, 2015) Despite the wide use and long history of anesthesia, the mechanism(s) of action that cause loss of consciousness and some of the side-effects, such as temporary amnesia, are not understood (Alkire, Hudetz and Tononi, 2008; Zurek *et al.*, 2014). The majority of general anesthetics are hydrophobic molecules, which dissolve well in lipid bilayers. A century ago, Meyer and Overton observed a positive lipid solubility and correlation between anesthetic potency. They hypothesized that a disturbance of the lipid bilayer was responsible for the loss of consciousness (LOC) (Gaus *et al.*, 2006; Lugli, Yost and Kindler, 2009). However, a large body of research failed to find a lipid-based mechanism, and at clinical concentrations general anesthetics minimally affect the biophysical properties of synthetic lipid bilayers (Herold *et al.*, 2017). Instead, it was discovered that some general anesthetics bind and potentiate γ-aminobutyric-acid-type (GABA) receptors directly (Bali and Akabas, 2004; Franks, 2008; Garcia, Kolesky and Jenkins, 2010). In addition, there is evidence that general anesthetics, including propofol, inhibit NMDA receptors (Sato *et al.*, 2005; Petrenko *et al.*, 2014), and modulate the activity of TWIK related $K^+$ (TREK-1) channels which have been found to be important for the loss of consciousness (Heurteaux *et al.*, 2004). However, for many anesthetics and channels no specific receptor binding sites have been found, leaving the possibility that perturbations of the local membrane environment are import for anesthesia.

Growing appreciation of functional nanodomains in the lipid bilayer of the cell membrane has revitalized efforts to study effects of anesthetics on membrane ultrastructure (Varma and Mayor, 1998; Simons and Toomre, 2000; Honigmann and Pralle, 2016; Sezgin *et al.*, 2017). Weak inter-molecular interactions cause lipids in the cell membrane to form transient nanodomains which in can phase separate into larger domains at lower temperature (Veatch and Keller, 2003; Levental *et al.*, 2009). It was found that a class of general anesthetics, N-alcohol anesthetics, lowers the critical temperatures at which lipid heterogeneities unmix in large unilaminar vesicles plasma membrane derived vesicles (Gray *et al.*, 2013). Also, super-resolution STORM microscopy of fixed cells showed that isoflurane perturbs cholesterol-rich nanodomains labeled by GM1 (Pavel *et al.*, 2020). However, it remains to be seen how these results translate to intact cells at physiological temperatures, when lipid nanodomains are smaller than the diffraction limit and protein association with them is very transient (Eggeling and Honigmann, 2014; Huang *et al.*, 2015). Here we present detailed measurements of the effect of propofol on three known types of lipid nanodomains in intact cells.

Evidence has been found that perturbations of lipid nanodomains by general anesthetics contribute functionally to the loss of consciousness. When hexadecanol, which stabilizes nanodomains in plasma membrane derived vesicles, is added to tadpoles under ethanol anesthesia, the depth of anesthesia is reduced (Machta *et al.*, 2016). Also, changes in the lipid nanodomain stability may modulate receptor and ion-channel activity. For example, it was reported that pentobarbital anesthesia reduced the lipid raft-association of GABA and NMDA receptors (Sierra-Valdez, Ruiz-Suárez and Delint-Ramirez, 2016). Recently, it was proposed that the effects of general anesthetics on TREK-1 are a results of a disruption of lipid nanodomains that cause activation of phospholipase D (PL-D) which modulates TREK-1 activity (Pavel *et al.*, 2020). Earlier work had demonstrated that local anesthetics lidocaine and chlorpromazine activate phospholipase C (PL-C) and decrease membrane-cytoskeleton adhesion (Raucher and Sheetz, 2001). This may explain how general and local anesthetics modulate TREK-1 activity in opposite directions (Heurteaux *et al.*, 2004; Pavel *et al.*, 2020). Although, Pavel et al had previously found that some local anesthetics, such as tetracaine and lidocaine, directly bind to the pore of TREK-1 and inhibit the channel (Pavel *et al.*, 2019)

Fundamentally, the evidence supports anesthetic-driven perturbation of phase separation in lipid bilayers. However, none of these studies directly quantified the effect of anesthetics on lipid nanodomains in intact cells at physiological temperature. Here, we present results of perturbations caused by propofol on three types lipid nanodomains in the external and internal leaflet of the plasma membrane of intact cells at physiological and at room temperature. Propofol's influence vastly differs at the two temperatures, changes over time, and strongly depends on the concentration across the clinically relevant range. The effects develop over a time-course of many minutes and are partially the result of active regulation by the cells actin cortex.

**Results**

**Quantification of Lipid Nanodomains in Outer and Inner Membrane Leaflet of Intact Cells**

To quantify lipid nanodomains in intact cells, we employed binned imaging fluorescent correlation spectroscopy (BimFCS) to quantify the diffusion of GFP tagged, nanodomain interacting membrane proteins on multiple length scales.(Sankaran *et al.*, 2010; Huang *et al.*, 2015; Jin, Simsek and Pralle, 2018) Binned imaging FCS uses a camera to acquire intensity data in single pixels and binned pixels simultaneously (**Fig. 1A,** see *Methods sections* for details). The temporal autocorrelation curves for the intensity data of each pixel and binned pixels are calculated and fitted with an appropriate diffusion model (**Fig. 1 B**).(Kraut, Bag and Wohland, 2012; Huang, Walker and Miller, 2015; Jin, Simsek and Pralle, 2018) The transit times $t_D$ through each detection area are plotted against the different area sizes, and fit with a straight line with y-offset to compute the intercept with the time axis for zero area, $t_0$ (**Fig. 1 C**). This $t_0$ is a measure of how strong the fluorescently tagged molecule is trapped by the nanodomains. It is the time that the molecule resided in a sub-100nm area in addition to the expected diffusion time through that area.

To establish a baseline for the strength of various nanodomain structures in intact cells, bimFCS was used to quantify the interaction of mGFP tagged markers with cholesterol nanodomains in the outer leaflet of the plasma membrane (using mGFP-GL-GPI), in the inner leaflet of the plasma membrane (using Lck10-mGFP), and with PI(4,5)P2 nanoclusters in the inner leaflet (using GFP-PLCδ-PH) (**Fig. 1D**). DiIC18, a lipid intercalating dye inert to nanodomains, served as a (cell membrane) control marker.

At physiological temperatures, bimFCS measures strong transient trapping of mGFP-GL-GPI in nanodomains in the outer leaflet ( $t_0$ = 10.0±2.5ms ), and slightly shorter transient trapping of GFP-PLCδ-PH and Lck10-mGFP in nanodomains in the inner leaflet ( $t_0$ = 6.7±1.9ms and $t_0$ = 5.0±2.1ms, respectively ) (**Fig. 1E**). These results are consistent with prior reports that all three markers interact with lipid nanodomains. The lipophilic dye DiIC18 gives $t_0$ of 0.3±0.2ms, verifying that it diffuses freely in the cell membrane without nanodomain interaction.

**Propofol Destabilizes Lipid Nanodomains in Intact Cells at Low Concentrations**

To study the effect of propofol on cholesterol nanodomains under physiological conditions, we quantified relative changes in nanodomains as function of time after propofol addition by acquiring data continuously on the same cell. PtK2 cells were incubated in a closed, temperature-controlled chamber on the microscope, and treated with two different concentrations of propofol, flanking the range of clinically relevant concentrations. In the clinic, propofol is administered as microemulsion intra-venously starting with a dose of 2 mg/kg body weight, then gradually increased until the desired depth of anesthesia is reached (Sahinovic, Struys and Absalom, 2018). In vivo, the actual effective molarity of propofol at the cellular level cannot be quantified and therefore cellular experiments are often performed at a range of concentrations from 5 µM to 100 µM (Sall *et al.*, 2012; Pavel *et al.*, 2020). For each cell $t_0$ is measured before the addition of propofol and again after 20 minutes of treatment, and reported as relative change $\Delta t_0/t_0$ (**Fig. 2**). After the treatment with 2µM, the $t_0$ values of all three separate nanodomain markers, mGFP-GL-GPI, GFP-PLCδ-PH and Lck10-mGFP, decreased by 25% after propofol treatments. These highly significant changes are comparable to those measured upon cholesterol oxidization (Huang *et al.*, 2015). The diffusion behavior of the control probe DiIC18 was not affected, demonstrating that the observed changes are specific to lipid nanodomains. The results demonstrate that at low concentrations propofol destabilizes cholesterol nanodomains in the outer and inner leaflet of the membrane. Similarly, the PIP2 nanoclusters in the inner leaflet were disrupted after 20 minutes 2 µMp. To examine whether the effect of propofol may be correlated to its anesthetic potency, we quantified the effect of propofol's structural analog 2,6-di-tert-butylphenol. 2,6-Di-tert-butylphenol has a ten-fold higher octanol/water partition coefficient than propofol, so will partition more into the membrane. However, even 100 μM of 2,6-di-tert-butylphenol neither potentiates GABA receptors nor induces loss of the righting reflex in tadpoles (Krasowski *et al.*, 2001). There was no detectable change of $t_0$ after 20 minutes 2 μM 2,6-di-tert-butylphenol treatment (**Fig. 2A**).

**Propofol Causes Stronger Lipid Nanodomains in Intact Cells at Higher Concentrations**

As several studies consider significantly higher concentrations of propofol to be to clinically relevant, another set of cells was incubated with 50 μM propofol for 20 minutes at 37ºC. After this treatment the $t_0$ values of mGFP-GL-GPI, GFP-PLCδ-PH and Lck10-mGFP *increased* significantly (**Fig. 2B**). The strongest increase, 163±28 %, was observed for mGFP-GL-GPI interacting with cholesterol stabilized nanodomains in the outer leaflet of the plasma membrane. In the inner leaflet the increases were smaller but still significant, 47±15 % for GFP-PLCδ-PH, and 25±6 % for Lck10-mGFP. The DiIC18 dye results show also at 50μM propofol no difference of $t_0$ at 50 μM. The effect of propofol on cholesterol nanodomains is further examined with its structural analog 2,6-di-tert-butylphenol, which is expected to strongly partition into the membrane but has no anesthetic potency. Even at 50μM 2,6-di-tert-butylphenol treatments do not modulate any of the three lipid nanodomains studied (**Fig. 2B**). An increase in $t_0$ indicates an increased nanodomain stability or size and is in stark contrast to the nanodomain destabilization by propofol at low concentrations. How does one membrane perturbing compound have opposite effects based on concentration? Prior work using plasma membrane derived vesicles only observe destabilization of lipid nanodomains over the entire concentration range (Gray *et al.*, 2013).

**Propofol Effect on Lipid Nanodomains in Intact Cells is Strongly Temperature Dependent**

To more narrowly identify the concentration of propofol at which its destabilizing activity counterbalances the stabilizing mechanism, we quantify the effects on the lipid nanodomains as a function of propofol concentration (**Fig. 3A**). At physiological temperature, propofol reduced the association of mGFP-GPI with lipid nanodomains for concentrations lower than 10 μM. At 10 μM stabilizing and destabilizing action cancel each other. Above 10 μM the cholesterol lipid nanodomains are stabilized by propofol. This effect slowly reached saturation above 100 μM propofol and has an $EC_{50}$ of about 60 μM (**Fig. 3A**).

In an attempt to silence most active processes in the cells, the experiments were repeated at room temperature. At 19°C propofol destabilized cholesterol lipid nanodomains in the membrane's outer leaflet at any concentrations between 1 μM and 150 μM. The amount of nanodomain destabilization saturates at 5 μM and no stabilization is observed. Therefore, we hypothesized that the nanodomain stabilization involves an active regularly mechanism that requires physiological temperatures in intact cells.

**The Destabilization of Lipid Nanodomains is Immediate, while Stabilization takes Minutes**

To identify possible regulatory mechanisms, we measured the time course of the propofol effect on lipid nanodomains for various propofol concentrations in intact cells at physiological temperature (**Fig. 6**). After two to five minutes of treatment, the earliest time interval accessible by bimFCS, any concentration propofol caused a reduction of the lipid nanodomain stability. At lower concentrations, 0.5 µM, 1.5 µM, and 2 µM, this destabilization persisted for the next 30 minutes (**Fig. 3B**). At higher concentrations, 10 µM, 20 µM, and 50 µM, nanodomains progressively stabilized (**Fig. 3C**). After about 20 minutes, the maximal nanodomains stabilization was reached for 10 µM and 20 µM propofol. For 50 µM propofol, the process completed only after 30 minutes.

**The Propofol Triggered Stabilization of Lipid Nanodomains Involves Actin Nucleation**

As the stabilization of the lipid nanodomains required about ten minutes and occurred only at physiological temperatures in intact cells, we hypothesized that it requires activity of the actin cortex forming the membrane cytoskeleton. It has been shown that active remodeling of cortical actin can regulate the spatiotemporal organization of cell surface molecules and affect the diffusion of GPI-anchored proteins(Gowrishankar *et al.*, 2012; Saha *et al.*, 2015). Therefore, we quantified the propofol effects on lipid nanodomains at 37°C in presence of inhibitors of either Myosin II motor activity, blebbistatin, or Arp2/3 actin nucleation activity, CK-666 (**Fig. 4A**). In control cells, 20 µM propofol increased $t_0$ by 60.7±16.6 %, but in cells treated with 4 µM blebbistatin, the increase significantly reduced to 38.7±4.8 %. The nanodomain stabilization is completely blocked by 4 µM CK-666, $\Delta t_0$ = 2.1±13.7 % (**Fig. 4A left**). At 50 µM propofol, near the $EC_{50}$ for nanodomain stabilization, Myosin II inhibition reduced the nanodomain stabilization by two thirds, $\Delta t_0$ = 46.7±14.8 % versus 154.3±12.4 % in control cells. Blocking Arp2/3 activity completely inhibits nanodomain stabilization and $t_0$ remains unchanged, $\Delta t_0$ = 0.6 ± 14.1 % (**Fig. 4A middle**). At saturating propofol concentration, 100 µM, the lipid nanodomains in control cells are fourfold more stable than without propofol $\Delta t_0$ = 408.6 ±42.6 %. Blocking Myosin II activity reduces this increase eightfold to $\Delta t_0$ = 56.4±16.5 %. Inhibiting Arp2/3 activity almost completely abolishes the increase, $\Delta t_0$ = 16.4±9.9 % (**Fig. 4A right**). The nanodomain stabilization was completely inhibited by blocking F-actin. In presence of 4 µM Latrunculin B, increasing amounts of propofol lead to increasing extends of nanodomain disruption (**Fig. S1**).

This data demonstrates that the nanodomains stabilization observed during higher propofol concentrations of longer incubations, is caused by activity of the underlaying cortical actin cytoskeleton. The largest contribution appears to be Arp2/3 mediated novel actin nucleation at the membrane, which is potentiated by Myosin II activity. To quantify directly changes in the meshwork of the membrane cytoskeleton, we used bimFCS to measure the hop-diffusion of a single transmembrane protein, mGFP-GT46 (see **Supplemental Information**). Low concentrations of propofol did not induce measurable changes in the membrane actin cytoskeleton in cells at physiological temperatures. However, at 50 μM propofol the number of corrals was significantly increased (**Fig. S2 A, B**), while the free diffusion within the corrals remained unchanged (**Fig. S2 C**). This data confirm that the local membrane properties remain unperturbed, but the number of cortical actin filaments near the membrane increased.

**Propofol Induced Stabilization of Lipid Nanodomains is Potentiated by GABA Receptor Signaling**

Propofol is known to potentiating and bind $GABA_A$ (γ-aminobutyric acid type A) receptors (Yip *et al.*, 2013). To determine whether GABA receptor activity contributes to the propofol induced stabilization of lipid nanodomains, we performed bimFCS measurements of the propofol effect on mGFP-GL-GPI in the presence of $GABA_A$ agonists and antagonists, at physiological temperature (**Fig. 4B,C**).

At 10 μM propofol the destabilizing and stabilizing effects on the lipid nanodomains balanced in control cells and there was no change, $\Delta t_0$ = 5.4±22.7. When under the same conditions, 10 μM GABA was added to the cells, the nanodomain stabilization was significantly increased, $\Delta t_0$ = 56.3±12.0 % (**Fig. 4B right**). The addition of GABA alone, without adding propofol, did not perturb the nanodomains, $\Delta t_0$ = 3.1±9.3 % (**Fig. 4B middle**). At 50 μM propofol, the lipid nanodomains in control cells were significantly stabilized, $\Delta t_0$ = 172.7±36.0 % (**Fig. 4C left**). Hence, we did not expect further potentiation by GABA, but instead added bicuculline to block the GABA receptor activity. In presence of 4 μM bicuculline and 50 μM propofol, the nanodomain stabilization was significantly decreased, $\Delta t_0$ = 76.1±4.9 % (**Fig. 4C right**). These data demonstrate that GABA receptor activation potentiates the nanodomain stabilization but is not necessary for the effect.

**Discussion**

The results show that the general anesthetic propofol affects the plasma membrane ultrastructure of living cells in a concentration, temperature, and time dependent manner. At room temperature, propofol disrupts lipid nanodomains in the plasma membrane at any concentrations.

At physiological temperatures however, propofol disrupts lipid nanodomains in the cell membrane of intact cells only at the lowest clinically relevant propofol concentrations and short incubation times. At higher, clinically relevant concentrations, and incubation times longer than ten minutes, propofol leads to a stabilization of lipid nanodomains in the plasma membrane as cells actively counter the initial disruption of the nanodomains. This nanodomain homeostasis involves ARP2/3 activity dependent novel actin nucleation ad the membrane and rearrangement of cortical actin by increased Myosin II motor activity. These regulatory effects are inhibited at room temperature. The signal for this response is at least in part the GABA receptor mediated calcium influx. Previous studies of effects of general anesthetics on lipid nanodomains in plasma membrane derived vesicles, model lipid bilayers or fixed cells failed to observe this active processes because it can only occur in living cells at 37°C.(Gray *et al.*, 2013; Herold *et al.*, 2017)

The discovered nanodomain homeostasis explains how a general anesthetic, which initially perturbs lipid nanodomains, eventually stabilizes these nanodomains beyond their original size. It provides the missing link in a recent study which associated isoflurane with increasing lipid nanodomains, activation of phospholipase D (PL-D) and potentiation of TREK-1 activity(Pavel *et al.*, 2020). Likely the changes in outer and inner leaflet nanodomains, including PIP2 cluster and dynamics of the cortical actin modulate other ion channels as well.

The data revealed an association between lipid nanodomain homeostasis in the membrane and cortical actin dynamics. This disruption of actin dynamics is likely at least in part responsible for the long-lasting memory defects found in some patients with general anesthesia. It remains to be seen whether in neurons the same relation holds. However, several studies have reported that general anesthetics can perturb actin at the synapse. Other have shown that volatile anesthetics block motility in dendritic spines,(Kaech, Brinkhaus and Matus, 1999) and that isoflurane anesthesia transiently perturbs the actin structure at the synapse.(Platholi *et al.*, 2014) A new preprint reports that propofol anesthesia in rats leads to week-long perturbation of actin structure and learning.(Zhang, Li and Wu, 2020) Those studies have suggested that the anesthetics act directly on actin, but found no evidence for a mechanism. Instead, our results suggest that the anesthetics primarily act in the lipid bilayer part of the membrane by destabilizing lipid nanodomains. The perturbed actin dynamics is then a consequence of cellular nanodomain homeostasis. In line with this, the prior studies only found actin dynamics near the membrane to be perturbed.

Stimulation of GABA receptors by propofol further potentiates the observed perturbation of actin dynamics and subsequent stabilization of lipid nanodomains by propofol. In general, calcium influx through the plasma membrane stabilizes inner and outer leaflet lipid nanodomains through electrostatic clustering of PIP2 and ARP2/3 mediated actin nucleation (Weixiang, unpublished data, (Jin, Huang and Pralle, 2014; Jin and Pralle, 2015)). These results support the idea that excitable cells contain a tight regulatory system between lipid nanodomains, electrostatic coupling of bilayer to cortical actin and channel signaling. General anesthetics interfere with some of the regulatory system. That interference is mostly transient, but in some cases can be long-lasting.

**Materials and Methods**

**Binned Imaging Fluorescence Correlation Spectroscopy (BimFCS)** uses total internal reflection fluorescence microscopy (TIRF) to illuminate the bottom membrane of cells grown on glass and a camera to acquire intensity data and treat pixels and binned pixels of the camera as observation areas(Sankaran *et al.*, 2010; Huang *et al.*, 2015; Jin, Simsek and Pralle, 2018) (**Fig. 1A**). Fluorescent tracers in the membrane are excited using a water-cooled Argon-Krypton laser, Innova 70C, Coherent), membrane of surface grown, intact cells, and a fast, high quantum yield, ultralow noise fiber coupled into an inverted microscope (AxioObserver, Zeiss) equipped with high NA objective lens (Zeiss, 100x oil, NA = 1.45]. Fluorescence data are collected by the objective, passed through emission filters (Semrock), and acquired by an EMCCD (Andor iXon+ 897) run at 600 frames per second. These data are binned according to the camera pixels into larger size n x n pixels, bleaching is corrected and the temporal autocorrelations for each pixel calculated (**Fig. 1 B**). When fitted with an appropriate diffusion model the transit time $t_D$ through each detection area size, and the number of fluorophores in each area are obtained(Kraut, Bag and Wohland, 2012; Huang, Walker and Miller, 2015; Jin, Simsek and Pralle, 2018). The transit times $t_D$ are plotted against the different area sizes, and fit with a straight line with y-offset. This intercept with the time axis for zero area, $t_0$, is a measure of how much the molecules are slowed down on the sub-100nm scale by interactions with domains. The reciprocal of the slope provide the effective diffusion constant $D_{eff}$ [details in(Jin, Simsek and Pralle, 2018)] in the same way as spot-variation FCS(Wawrezinieck *et al.*, 2005), but with advantage of only requiring one single measurement. The analysis is performed using custom code written in IgorPro (Wavemetrics).

**Cells and Protein Constructs, and Chemicals**. Experiments where performed in male long-nosed potoroo epithelial kidney (PtK2) cells (NBL-5 from ATCC Manassas, VA). Cells were plated on Poly-L-Lysine (Sigma Aldrich) coated glass coverslips (Carolina Biological). The next day, cells are transfected using Lipofectamine 3000 (Thermo Fisher Scientific), with plasmids encoding either mGFP-GL-GPI(Keller *et al.*, 2001), Lck10-mGFP, or GFP-PLCδ-PH. mGFP-GL-GPI was made by introducing the A206K(Zacharias *et al.*, 2002) mutation in eGFP-GL-GPI which was a gift from Patrick Keller(Keller *et al.*, 2001). Similarly, Lck10-mGFP was made from Lck10-eGFP which was gifted by Ron Vale(Douglass and Vale, 2005). GFP-PLC$\delta$-PH a gift from Tobias Meyer (Addgene #21179)(Stauffer, Ahn and Meyer, 1998).

To acquire bimFCS data, the coverslips with cells are washed and transferred into a custom-built sample holder inside an incubation chamber on the microscope stage. The cells are imaged in physiological salt solution (PSS, ingredients in mM: CaCl2 2, NaCl 151, MgCl2 1, KCl 5, HEPES 10, Glucose 10 pH 7.3).

Anesthetics and chemicals were purchased purified and diluted in PSS on the day of the experiment: propofol solution (1.0 mg/mL in methanol); 2,6-di-tert-butylphenol; $\gamma$-aminobutyric acid (GABA); bicuculline; blebbistatin; CK-666 (all Sigma Aldrich); and latruncilin B (Cayman Chemical).

## Acknowledgments

The authors acknowledge support from the Department of Physics and the University at Buffalo, and thank Heng Huang and Muhammed F. Simsek for advice on the bimFCS technique, and Sara Parker and Jason Myers for technical molecular biology support.

**Figures and Tables**

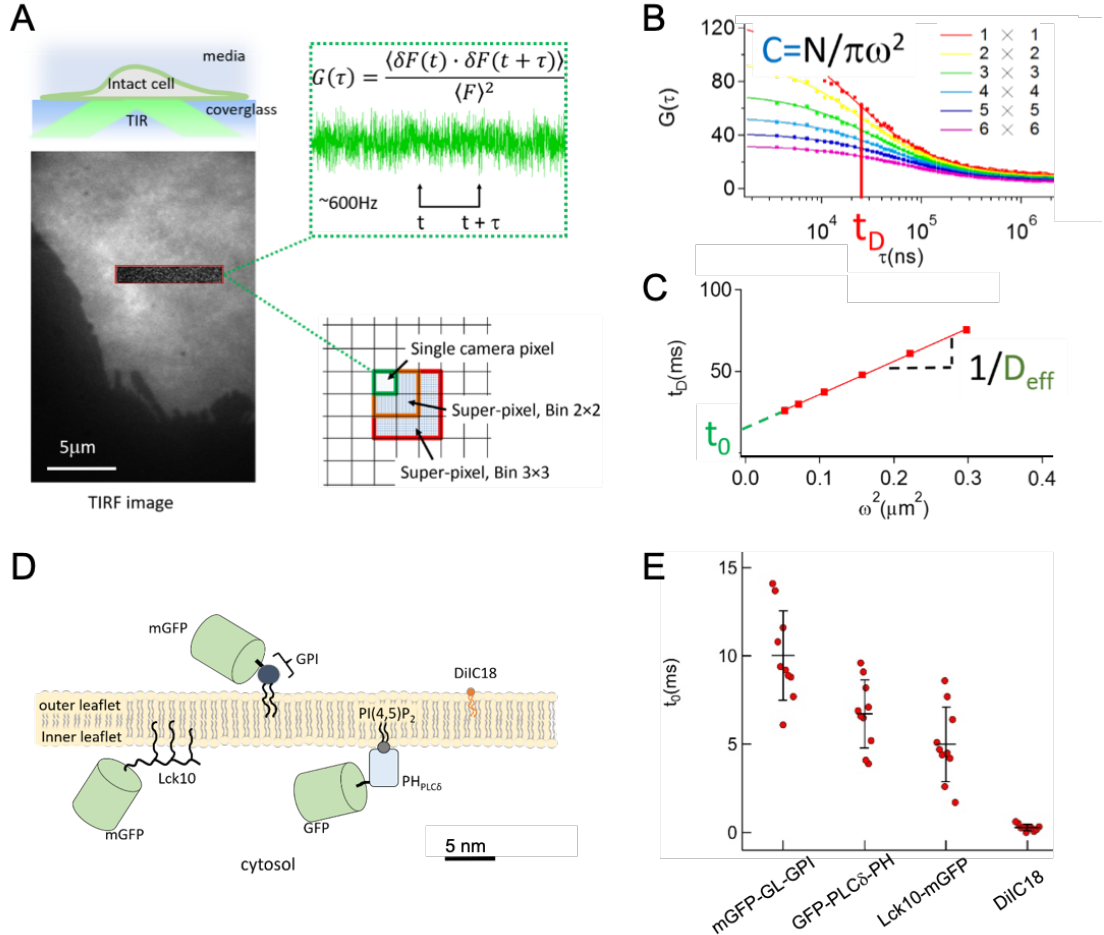

**Figure 1. Principle of bimFCS analysis.** (**A**) TIRF illumination excites fluorophores in the bottom membrane of intact cells on a glass. An EMCCD camera collects the fluorescence intensity, then the temporal autocorrelation is computed for each pixel as well as for a series of binned camera pixel. (**B**) The fit of the autocorrelation curves provides a transit time $t_D$ and number of fluorophores for each pixel and binned pixel. (**C**) The transit times $t_D$ are plotted against the differently sized detection areas $\omega^2$ and fit it with a straight line. This graph allows to extract an intercept with the time axis for zero area, $t_0$, and an effective diffusion constant, $D_{eff}$, the reverse of the slope. (**D**) Schematic drawing of membrane fluorescent markers used in this study: mGFP-GL-GPI as probe for cholesterol stabilized nanodomains in the outer leaflet of the plasma membrane; Lck10-mGFP as probe for cholesterol stabilized lipid nanodomains marker in the inner leaflet of the plasma membrane; and GFP-PLCδ-PH as probe for electrostatically stabilized PI(4,5)P2 domains in the inner leaflet. marker; the freely diffusing DiIC18 lipid dye was used as control. (**E**) Summary of baseline $t_0$ values of these markers in PtK2 cells at physiological temperatures.

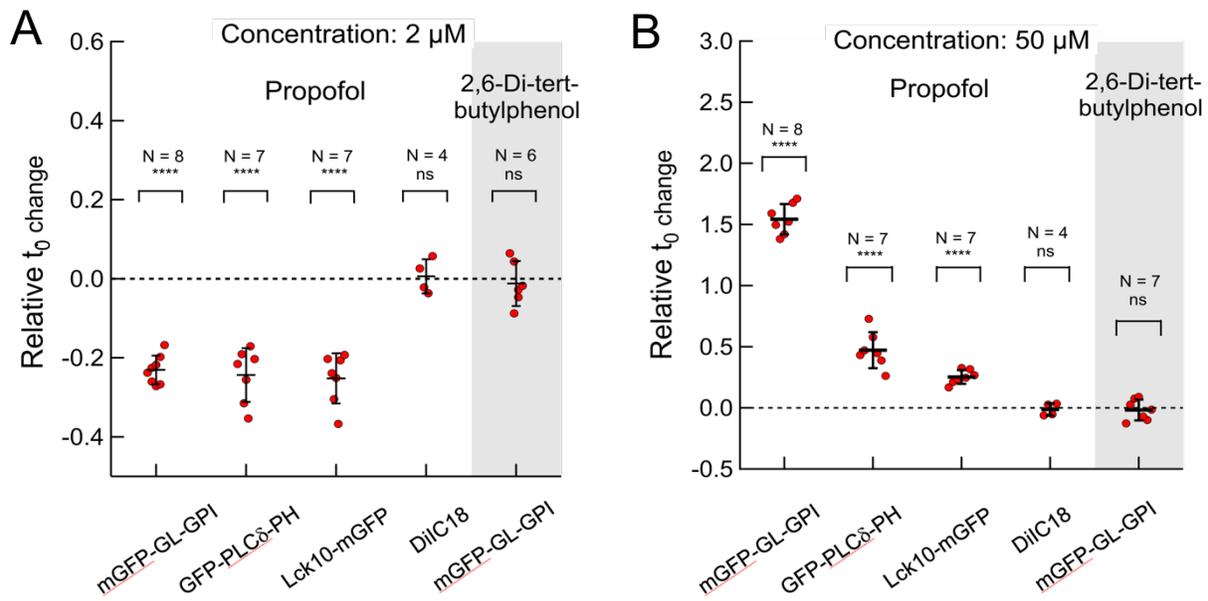

**Figure 2. Propofol induced change of the association of various mGFP tagged markers with lipid nanodomains. (A)** Effects of 2 μM propofol or the non-anesthetic analog 2,6-di-tert-butylphenol: For propofol the nanodomains measured by mGFP-GL-GPI, by GFP-PLCδ-PH, and by Lck10-mGFP were all reduced by about 25%, which is highly significant ( $p \leq 0.0001$ ). The value for DiIC18 (N = 4) remains unchanged. At 2μM 2,6-di-tert-butylphenol, the nanodomains measured by mGFP-GL-GPI remained unchanged. (**B**) Effects of 50 μM propofol or the non-anesthetic analog 2,6-di-tert-butylphenol: At 50 μM propofol, the nanodomain stability of domains measured by mGFP-GL-GPI increased 160 %, while PIP2 domains in the inner leaflet measured by GFP-PLCδ-PH increased by 50%, and the ones interacting with of Lck10-mGFP increase by a quarter ( all highly significant, $p \leq 0.0001$ ). Even at high propofol concentrations, the diffusion of DiIC18 remained unchanged. Similarly, 50 μM 2,6-di-tert-butylphenol did not perturb lipid nanodomains as measure use mGFP-GL-GPI.

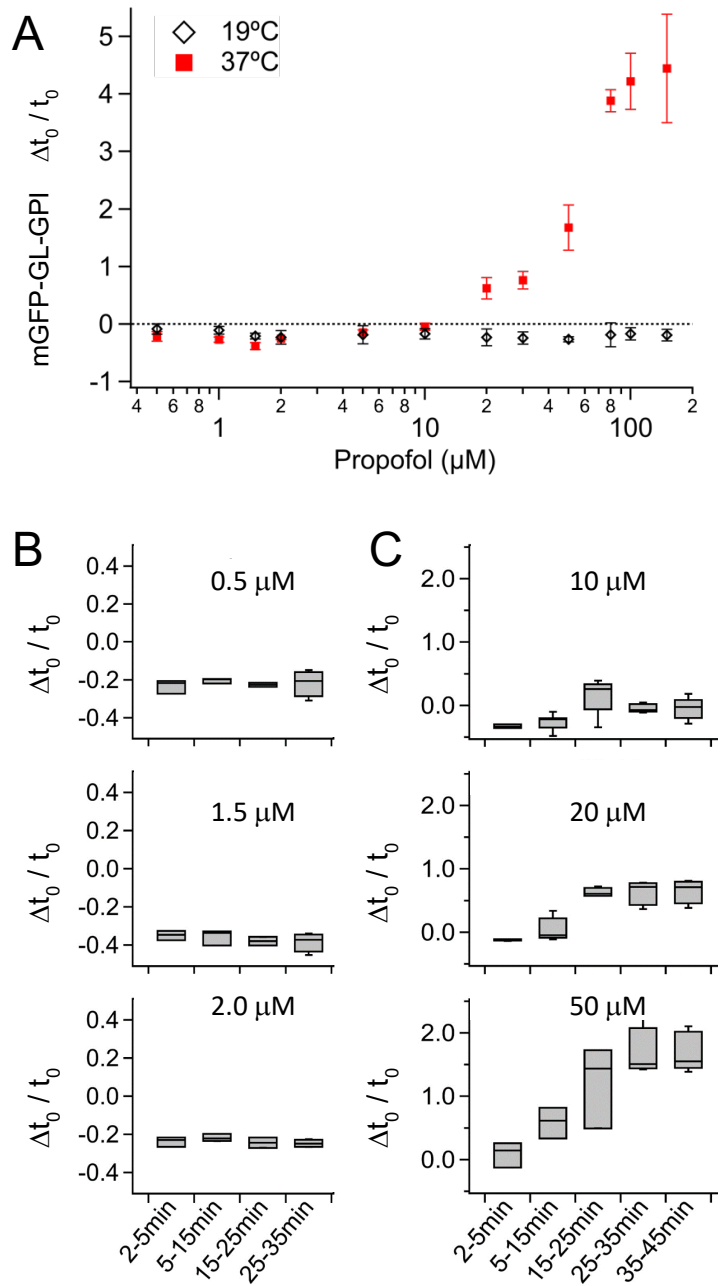

**Figure 3. (A) Propofol modulation of mGFP-GL-GPI association with lipid nanodomains in cells are function of propofol concentration and differ widely between 19°C and 37°C.** At 19°C, all propofol concentrations resulted in a reduction lipid nanodomains by 20% ( gray diamonds ). At 37°C, low propofol concentrations destabilized the nanodomains but larger ones strongly stabilized nanodomains ( red squares ) ( n > 20 for each condition; Mean ± SEM ). **(B,C) Time dependence of propofol effect on mGFP-GL-GPI.** The nanodomain association of mGFP-GL-GPI was measured repeatedly over 35 minutes in PtK2 cells incubated with various propofol concentrations at 37°C.

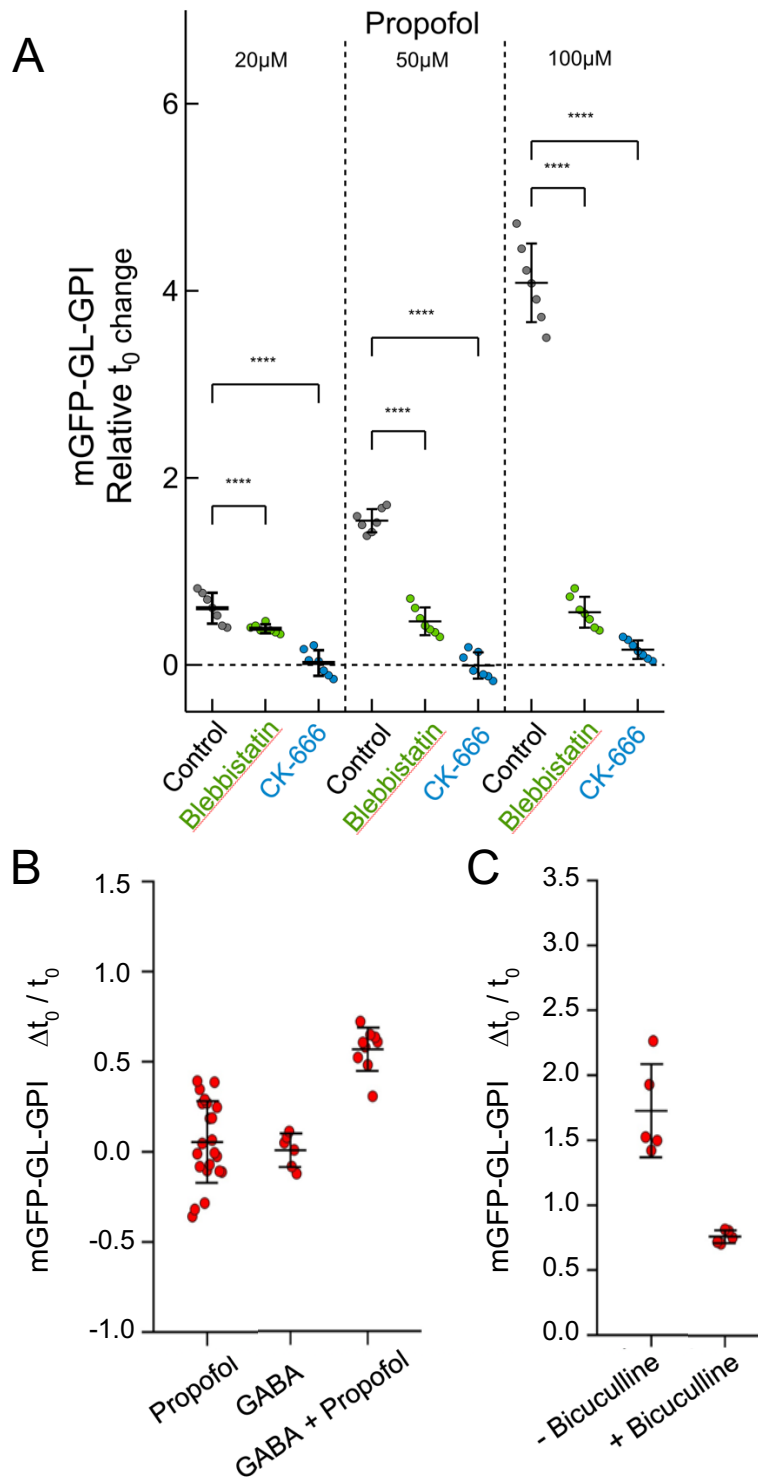

**Figure 4. (A) Inhibitors of cortical actin dynamics reduce propofol induced stabilization of lipid nanodomains**. For three propofol concentrations, 20 μM, 50 μM, 100 μM (left to right), we measured the change in nanodomain stability in control cells at 37°C (black), in cells pretreated

with 4 µM blebbistatin (green) and cells pretreated with 4 µM CK-666 (blue). For all propofol concentrations blebbistatin significantly reduced and CK-666 abolished the nanodomain stabilization. **(B) GABA receptor activation potentiates propofol induced nanodomain stabilization.** Lipid nanodomain stabilization measured as change of $t_0$ for mGFP-GL-GPI when cells were treated either with 10 µM propofol or 10 µM GABA alone, or with both 10 µM GABA and 10µM propofol together. **(C) Blocking GABA receptor activity reduces propofol induced nanodomain stabilization.** Quantification of nanodomain stabilization by 50 µM propofol addition in the absence (-) and presence (+) of the GABA receptor inhibitor bicuculline (4 µM).

# Appendix: Supplemental Methods and Controls

**Quantifying Changes of the Membrane Cortical Actin Network by Measuring Hop-diffusion**

Diffusing transmembrane proteins with significant cytoplasmic tail but no binding to other proteins encounter filaments of the underlying cytoskeleton as steric hindrance causing corralled or hop-diffusion (Kusumi and Sako, 1996; Kusumi and Suzuki, 2005; Kusumi *et al.*, 2011). In combination with a GFP tagged single transmembrane domain protein as probe, svFCS and bimFCS can quantify properties of the membrane cytoskeleton close to the lipid bilayer (Wawrezinieck *et al.*, 2004, 2005; Jin, Simsek and Pralle, 2018). Hop diffusion consists of free Brownian type diffusion on short length and time-scale, but on longer time-scales a greatly reduced diffusion due to the low hopping possibility to pass through cytoskeleton meshwork. As a result, FCS curves are interpreted as temporal autocorrelation from two populations of molecules with two separate diffusion coefficients: one for the free diffusion $D_f$ within corrals and one for hop-diffusion $D_{hop}$ (Details in Jin et al. 2018). The confinement strength S is the ratio of the two diffusion coefficients $D_f / D_{hop}$. It quantifies how much the corrals confine the diffusion of the transmembrane proteins. The length L measures the average edge length of the cytoskeletal corrals.

**Figure S1** shows that the observed stabilizeation of nanodomains at higher propofol concentraitons require an intact actin cytoskeleong as latrunculin B treatment abolishes lipid nanodomains stabilization in the cells.

**Figure S2** shows relative changes of a corral size ( $L / L_o$ ), strength of confinement ( $S / S_o$ ) and free diffusion ( $Df / Df_o$ ). Low concentrations of Propofol do not induce measurable changes in the membrane actin cytoskeleton in intact PtK2 cell at physiological temperatures. At 50 µM

Propofol increase the number of corrals as apparent in a 40% reduced average corral edge length L (**Fig. S2 A**) and in a 40% increased confinement strength S (**Fig. S2 B**). The free diffusion of the transmembrane probe within the corral remains unchanged (**Fig. S2 C**), consistent with result that the local membrane properties remain unperturbed.

**Figure S3** shows that 30 minutes of profol treatment does not affect subsequent cell growth, ruling out that the observed effects are a consequence of reduced cell health.

## References for appendix

# Appendix: Supplemental Figures

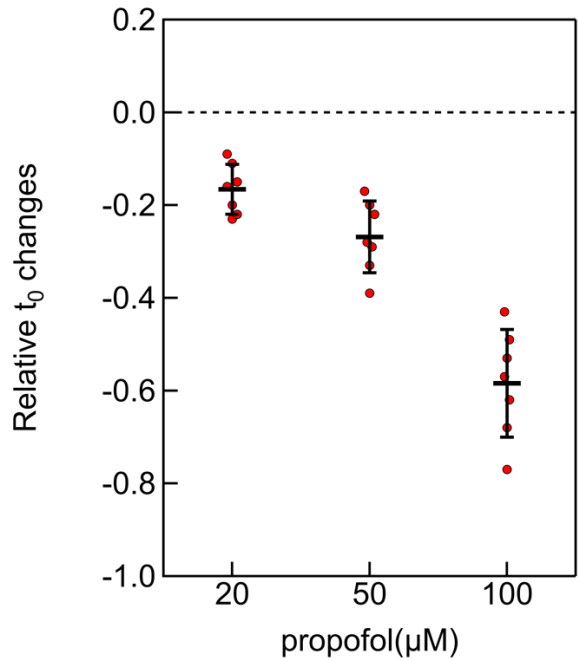

**Fig. S1**. **Latrunculin B treatment abolishes lipid nanodomains stabilization in intact cells.** The effect of various propofol concentration on the strength of the interaction of mGFP-GL-GPI with cholesterol nanodomains was quantified in PtK2 cells (at 37°C). In the presence of 4 μM Latrunculin B increasing propofol concentration lead to increasingly larger disruption of the lipid nanodomains.

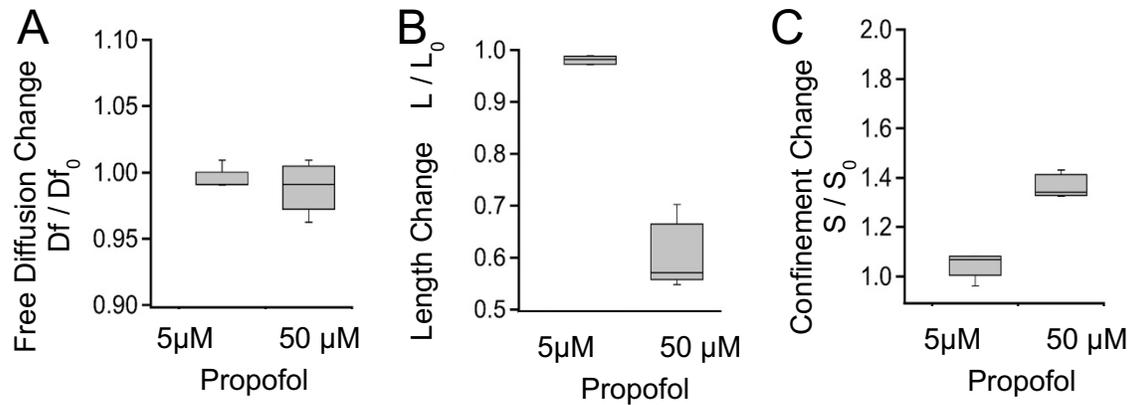

**Fig. S2**. **Propofol does not affect free diffusion of non-raft transmembrane protein but at higher concentrations perturbs of the cortical actin network close to the lipid bilayer.** BimFCS analysis of hop-diffusion of mGFP-GT46 in PtK2 cells at physiological temperatures. The effects are quantified as change of local free diffusion within actin corrals, edge length of actin corrals and strength of confinement. (**A**) Neither 5 μM nor 50 μM propofol affect the free diffusion of the transmembrane protein mGFP-GT46 locally, within actin corrals ( $D/D_0$= 0.99±0.02 (n=7) and 0.98±0.03 (n=8)). (**B**) Propofol at 5 μM did not affect the corral edge length ( $L/L_0$ = 0.98±0.02 (n=7)). However, at 50 μM propofol L is reduced significantly to 0.62±0.07 (n=8 cells). (**C**) Similarly, the confinement within the corrals is not significantly affected by 5 μM propofol ( $S/S_0$ = 1.07±0.07 ), but 5 μM propofol increase the confinement significantly ( $S/S_0$ = 1.36±0.08 ).

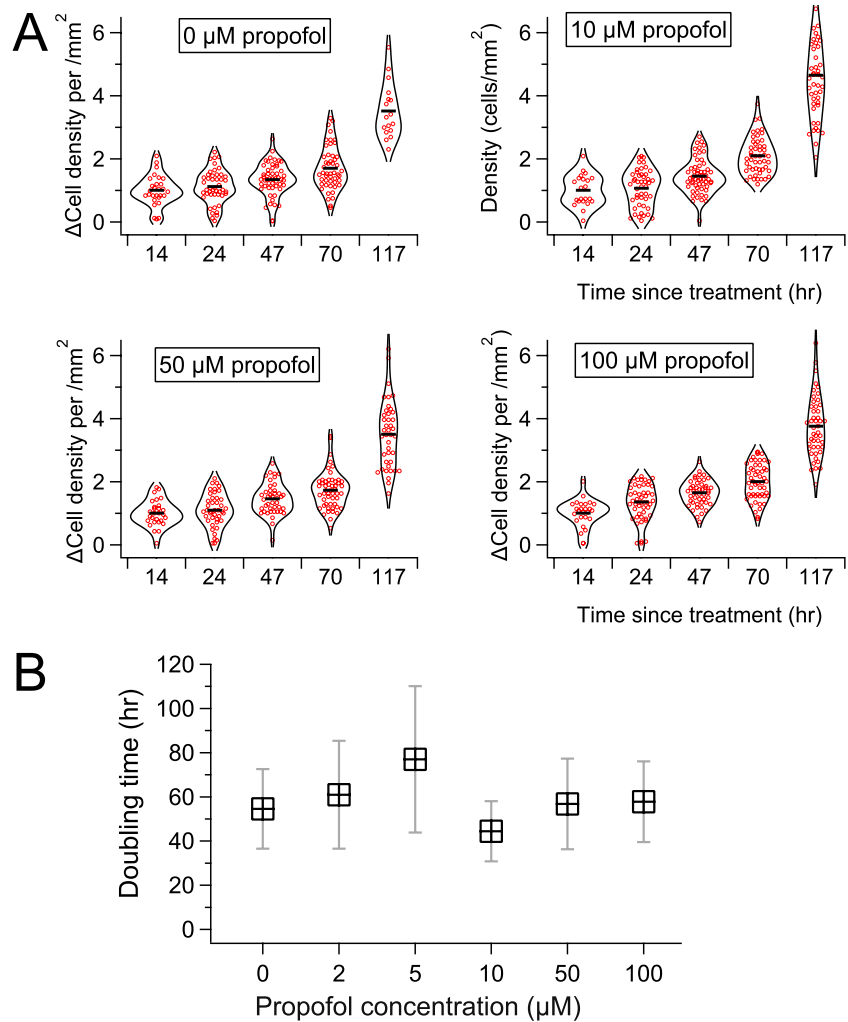

**Fig. S3. Incubation of PtK2 cells in PSS buffer with propofol did not affect subsequent growth.** (**A**) Cells were incubated for 30 min at 37°C in the indicated concentrations of propofol. Then, the cells were returned to growth media and placed in the incubator. Cell density was counted at six time points after the propofol treatment. Propofol concentrations of 0 μM, 2 μM, 5 μM, 10 μM, 50 μM, 100 μM were tested. (**B**) A growth curve was fitted to each data set of cell densities, and the doubling time for each propofol treatment calculated. There was no significant difference of the doubling time between the five propofol concentrations and the control buffer solution.